# Coalescence-induced late departure of bubbles improves water electrolysis efficiency

Tao Wu[1,3], Bo Liu*[1,3], Haohao Hao[2], Fang Yuan[1], Yu Zhang[1], Huanshu Tan*[2], Qiang Yang*[1]

[1] Department of Mechanical and Power Engineering, East China University of Science and Technology, Shanghai, 200237, China;

[2]Multicomponent Fluids group, Center for Complex Flows and Soft Matter Research & Department of Mechanics and Aerospace Engineering, Southern University of Science and Technology, Shenzhen, 518055, Guangdong, China.

[3]These authors contributed equally: Tao Wu, Bo Liu.

*Corresponding author. E-mail address: boliu@ecust.edu.cn ; qyang@ecust.edu.cn

tanhs@sustech.edu.cn

**Abstract**

In water electrolysis, bubbles form on the electrode and interact through processes such as collision and coalescence. However, the impact of bubble coalescence—a fundamental process governing electrolytic bubble behaviour—on electrolysis efficiency remains unclear. Here, we show that enhancing bubble coalescence improves electrolysis efficiency by more than 30% compared to systems where coalescence is inhibited. One key feature is the continuous coalescence of a newly detached bubble with microbubbles on the electrode, which delays the former from departing. Experimental observations and numerical simulations reveal two key benefits of bubble coalescence for electrolysis efficiency: (1) it liberates surface bubbles from the electrode at much smaller sizes, reducing their diameter from approximately 60-80 μm to less than 10 μm, thus freeing the active sites of the electrode from bubble coverage; (2) it induces strong agitation, with velocities reaching ∼1m/s in a small region near the electrode (at a depth of $10^{-5}$ m), thereby significantly improving the heat/mass transfer locally. Importantly, the chaotic agitation effect lasts for approximately 10 ms, two orders of magnitude longer than the coalescence process, which occurs in around 0.2 ms. This work provides valuable insight into bubble management in water electrolysis and other gas-evolution electrochemical reactions.

**Introduction**

The production of green hydrogen via the hydrogen evolution reaction (HER) in water electrolysis is anticipated to be crucial in achieving global "net-zero emission" targets by reducing carbon emissions in the chemical engineering, transportation, and steel industries.[1,2] Enhancing electrolysis efficiency to lower costs is essential for large-scale green hydrogen production through water electrolysis. During water electrolysis, gas bubbles form on the electrode surface, which significantly reduces efficiency by blocking active sites and hindering ion conduction pathways (Fig. 1a, left) [3]. Thus, minimizing the effect of bubbles during water electrolysis is crucial for improving its efficiency.

The detachment of electrolytic bubbles from the electrode depends on both bubble-electrode and bubble-bubble interactions. Most research has focused on bubble-electrode interactions, demonstrating that a reduction in bubble departure size enhances electrolysis efficiency[4–6]. For instance, Koza et al. found that applying a magnetic field reduced the bubble detach size from 0.6 mm to 0.4 mm, increasing current density by nearly 40% at the same HER potential[7]. As a result, methods such as modifying electrode structure/materials[8–11] and applying external physical fields[12–14] have been developed to reduce bubble detachment size and enhance improve efficiency.

However, despite the critical role of bubble collision and coalescence in bubble departure from the electrode[15–17], the impact of bubble-bubble interactions on electrolysis efficiency remains largely unexplored. Colliding bubbles generate additional forces due to liquid drainage[18–20] and surface energy release upon coalescence[21–23], significantly affecting the bubble detachment. For example, the coalescence of surface bubbles can trigger self-propelled detachment[22–26], and already detached bubbles may increase in size by merging with surface bubbles still growing on the electrode[27]. Recent research indicates that modulating bubble coalescence affects both bubble departure size and local mass transport[26,28,29]. Therefore, understanding how bubble-bubble interactions influence electrolysis efficiency is expected to open new avenues for technological innovation.

In this work, we conducted microelectrode experiments to explore the impact of bubble-bubble interactions on the HER efficiency. We systematically reduced the coalescence probability of colliding bubbles by adding varying species and concentrations of electrolytes, such as $HClO_4$, $Na_2SO_4$, into a 0.5 M $H_2SO_4$ solution. Surprisingly, we observed a distinct decrease in HER efficiency by around 30%, accompanied by a 50% reduction in bubble departure diameter upon the addition of $HClO_4$. This finding challenges the conventional understanding that reducing the departure diameters improves electrolysis efficiency, a perspective largely informed by studies focusing on bubble-electrode interactions. By analysing the bubble departure process, we propose that the rapid coalescence of surface bubbles and the just-detached bubble enhances the detachment of the surface bubbles and induces strong agitation at the electrode surface, thereby enhancing electrolyte-electrode interface heat and mass transfer, ultimately promoting electrolysis efficiency. This beneficial effect is diminished when bubble coalescence is inhibited.

## Results and Discussion

### Effect of bubble bubble-bubble interaction on electrolysis efficiency

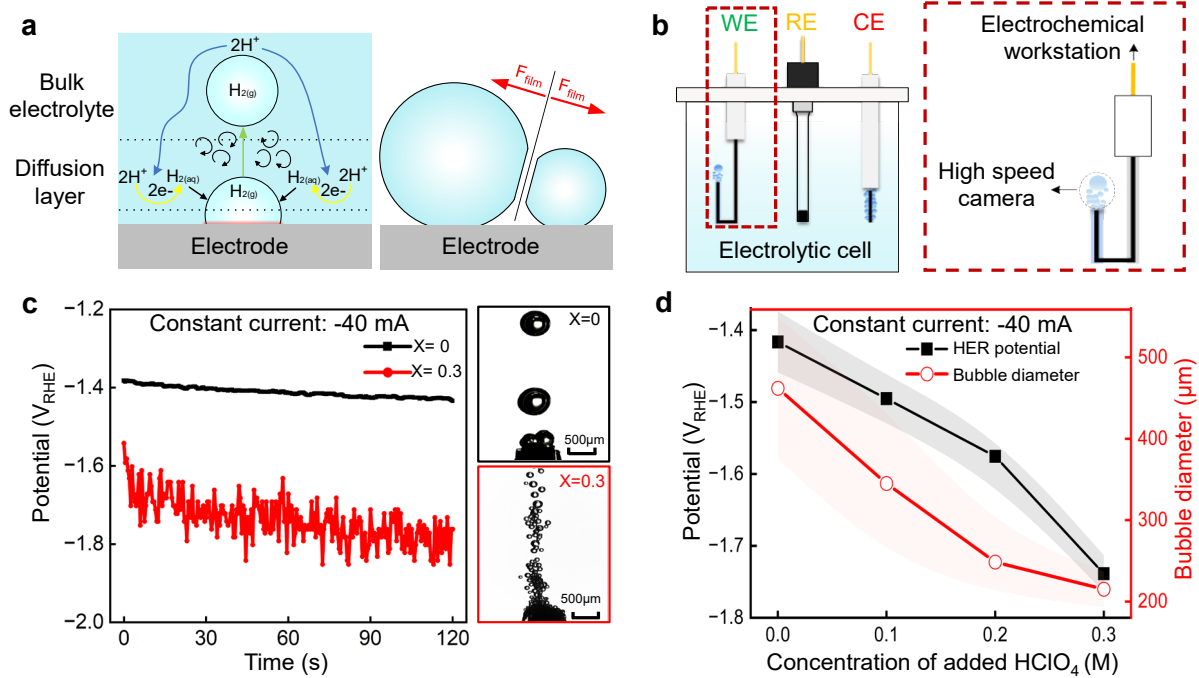

**Figure 1. | Illustration of bubble-bubble interaction and its impact on water electrolysis. a,** Schematic illustrations of (left) bubble coverage on the electrode surface and blockage of the diffusion pathway, where the black curved arrows indicate the local turbulence induced by detaching bubbles (adapted from ref.[3] with permission), and (right) the repulsion force ($F_{film}$) formed from film drainage between two colliding bubbles. **b,** Schematic diagram of the three-electrode experimental system, where a Pt microelectrode (diameter $\Phi$ =1mm) is used as the working electrode (WE), Hg/Hg$_2$SO$_4$ reference electrode (RE), and a Pt plate (10 mm × 10 mm × 0.2 mm) counter electrode (CE). **c,** The HER potential during constant current electrolysis at -40 mA (with a current density of approximately 5 A/cm²) and snapshots of electrolytic bubbles in 0.5 M H$_2$SO$_4$ + X M HClO$_4$ (X = 0, 0.3) solutions. **d,** Relationship between the average bubble detachment diameter and HER potential at different HClO$_4$ concentrations (X=0, 0.1, 0.2, 0.3) at -40 mA, the shaded error bars representing standard deviation.

The experiment utilized a three-electrode electrolytic cell (Fig. 1b) with a custom U-shaped platinum wire (99.9% purity, 1 mm diameter) as the working electrode. This electrode was enclosed in a glass capillary tube, allowing only the top cross-section to contact the electrolyte. A unique feature of this microelectrode setup is that bubble growth primarily occurs via coalescence[16], providing a controlled environment to study bubble interactions on electrolysis efficiency. An electrochemical workstation (Corrtest CS1350) recorded changes in HER potential, while a high-speed camera (Photron Mini AX200) captured hydrogen bubble characteristics (see supplementary information for details). Electrolysis efficiency is evaluated by HER potential; a more negative potential indicates higher energy consumption at constant current and, thus, lower efficiency[30].

The addition of 0.3 M $HClO_4$ to 0.5 M $H_2SO_4$ yields surprising results (Fig. 1c). With the addition of $HClO_4$, the electrolyte conductivity increases[31,32] and bubble departure size decreases[4,6,33], both are expected to improve electrolysis efficiency. However, the experimental results show the opposite (see Fig.1 c); that is, the addition of $HClO_4$ leads to an approximately 20% sharp decrease (HER potential is more negative at the same current) in electrolysis efficiency (Fig. 1c), as the HER potential shifts from -1.42 V to -1.74 V at a current of -40 mA. The decrease in electrolysis efficiency with smaller bubble departure sizes is observed across various concentrations of $HClO_4$ (see Fig. 1d) and is also observed in systems using larger plate electrodes (see Fig. S3).

We hypothesize that the unexpected result stems from variations in bubble coalescence behaviour across different solutions. While bubbles coalesce rapidly upon collision in 0.5 M $H_2SO_4$, this process is markedly suppressed with the addition of $HClO_4$ (see Fig. 1c and Supplementary Videos 1-2). Compared to conductivity, bubble coalescence may play a dominant role in determining electrolysis efficiency. Moreover, the conventional belief that smaller bubbles enhance efficiency—primarily based on bubble-electrode interactions—may not hold when bubble-bubble interactions become more influential. To elucidate the underlying mechanisms, we systematically investigate how factors influencing bubble coalescence[20,34,35], such as electrolysis current, electrolyte type, and concentration, affect electrolysis efficiency.

**Effect of electrolysis current on bubble coalescence and HER potential**

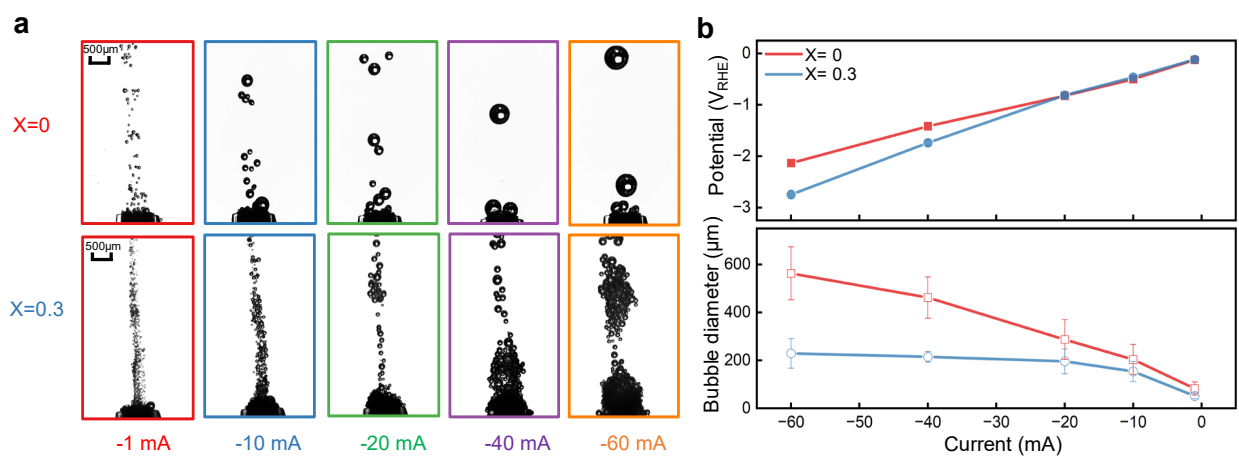

**Figure 2. | Effect of electrolysis current on bubble coalescence and HER potential. a,** Snapshots of electrolytic bubbles produced in 0.5 M $H_2SO_4$ and 0.5 M $H_2SO_4$ +0.3 M $HClO_4$ solutions at different currents. **b,** Variations in HER potential (above) and bubble departure diameter (below) with current in 0.5 M $H_2SO_4$ and 0.5 M $H_2SO_4$ +0.3 M $HClO_4$

The first factor considered is the electrolysis current, which correlates linearly with gas production rate, therefore increasing the frequency of bubble collisions and coalescence in the system. Two solutions, with and without 0.3 M $HClO_4$, differ in their probability of bubble coalescence as noted previously[34,36,37]. Fig. 2 shows the experimental results across a wide electrolysis current from -1 mA to -60 mA. Starting with a similar bubble departure size of around 60-80 μm in both solutions at -1 mA, the bubble size gradually increases to 560 μm in 0.5 M $H_2SO_4$, and to 230 μm with the addition 0.3 M $HClO_4$ when the current reaches -60 mA. The increase in bubble departure size with current can be attributed to the improved bubble coalescence resulted from the increased frequency of bubble collisions, while the smaller bubble departure size with the addition of $HClO_4$ can be explained by the reduced bubble coalescence probability.

HER potential (efficiency) is correlated closely with the difference in bubble departure size, with higher efficiency corresponding to larger bubble departure sizes. The HER potential is nearly identical in the two solutions at a small constant current of -1 mA, but a gap emerges and widens as the current increases, reaching approximately 30% at -60 mA (Fig. 2b above). This observation strongly reinforces the expectation that the inhibitory effect of bubble coalescence on bubble size and HER efficiency becomes even more significant at higher currents, where gas production rates are greater.

**Effects of electrolyte species and concentration on bubble coalescence and HER potential.**

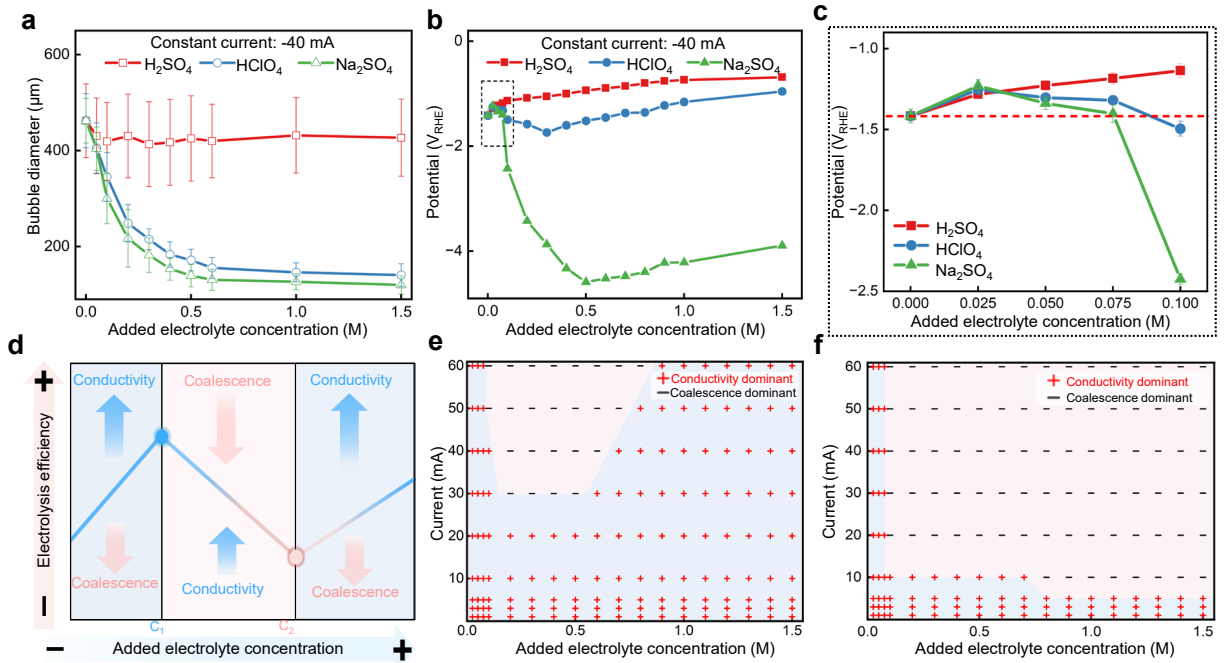

**Figure 3. | Effect of electrolyte concentration and species on bubble coalescence and HER potential.**
**a,** Bubble departure diameter in 0.5 M $H_2SO_4$ with varying concentrations of $H_2SO_4$, $HClO_4$, and $Na_2SO_4$. While both $HClO_4$ and $Na_2SO_4$ inhibit bubble coalescence, $H_2SO_4$ does not inhibit bubble coalescence. **b,** HER potential variation with different electrolytes concentrations of $H_2SO_4$, $HClO_4$ or $Na_2SO_4$. **c,** Local magnification of the dashed region in Figure 3b corresponding to the concentration range 0-0.1 M. **d,** Schematic of the dominant mechanism (coalescence inhibition or increased conductivity) affecting the electrolysis efficiency at high current, with transition concentrations $c_1$ and $c_2$. For $HClO_4$, $c_1 \approx 0.025$ M, $c_2 \approx 0.3$ M; for $Na_2SO_4$, $c_1 \approx 0.025$ M, $c_2 \approx 0.5$ M. **e** and **f,** Diagrams showing the impact of current and concentration on electrolysis efficiency in 0.5 M $H_2SO_4$ + X M $HClO_4$ (e) and 0.5 M $H_2SO_4$ + X M $Na_2SO_4$ (f). Positive $\Delta E$ (defined as $E_{(0.5\ M\ H_2SO_4 + X\ M\ HClO_4/Na_2SO_4)} - E_{(0.5\ M\ H_2SO_4)}$) represented by red +, suggest the electrolysis efficiency with added electrolyte is higher than the 0.5 M $H_2SO_4$ solution; whereas a negative $\Delta E$, represented by black -, suggest a lower efficiency.

The second factor explored is the type of electrolytes ($H_2SO_4$, $HClO_4$, and $Na_2SO_4$) added to 0.5 M $H_2SO_4$. Notably, the addition of $H_2SO_4$ does not inhibit bubble coalescence, while both $HClO_4$ and $Na_2SO_4$ significantly suppress it[38]. According to Fig.3a, increasing the concentration of $H_2SO_4$ to 1.5 M does not significantly change the bubble departure diameter, which remains between approximately 460 and 420 μm. In contrast, both $HClO_4$ and $Na_2SO_4$ significantly reduce the bubble departure diameter with increasing concentration, decreasing it from approximately 460 μm to less than 200 μm at 0.5 M. This reduction, indicating inhibited bubble coalescence, plateaus at concentrations greater than 0.5 M. The bubble size in $Na_2SO_4$ solution is smaller than in $HClO_4$ solution at the same concentration, which agrees with the expectation that $Na_2SO_4$ has a stronger ability to inhibit bubble coalescence than $HClO_4$[36], further confirming that bubble coalescence is critical for bubble departure size.

As expected, we observed a clear correlation between the type of support electrolyte and electrolysis efficiency, as depicted in Figs. 3b and c. $H_2SO_4$ shows a marginal increase in efficiency with increasing concentration. This occurs because the addition of $H_2SO_4$ does not inhibit bubble coalescence but can improve solution conductivity. In contrast, $HClO_4$ initially decreases the efficiency at concentrations of up to 0.3 M, after which the efficiency increases at higher concentrations. $Na_2SO_4$ drastically reduces the HER potential by about 200%, from -1.42 V to -4.59 V with increasing concentration from 0 to 1.5 M, highlighting its strong negative impact on efficiency. A possible explanation for this sharp decrease is that $Na_2SO_4$ not only inhibits bubble coalescence but also provides $Na^+$, which competes with $H^+$ ions at the reaction interface[39,40] and further decreases the electrolysis efficiency. Overall, the experimental results show that the stronger the electrolyte's ability to inhibit bubble coalescence, the greater its impact on bubble size and electrolysis efficiency; that is, $Na_2SO_4$>$HClO_4$>$H_2SO_4$.

The third factor explored is the concentration of electrolyte added to 0.5 M $H_2SO_4$ due to its impact on coalescence inhibition[36,38,41]. Even for coalescence inhibition electrolytes such as $HClO_4$, a critical concentration of approximately 50-100 mM is required for the inhibition effect to manifest[38]. The inhibition effect increases with electrolyte concentration but plateaus beyond a certain concentration, typically approximately 500 mM, according to bubble coalescence research[34].

The variation of added electrolyte concentration helps to uncover the competitive effects of conductivity and coalescence inhibition on the electrolysis efficiency. The impact of electrolytes such as $HClO_4$ and $Na_2SO_4$ on inhibiting bubble coalescence becomes significant only after surpassing the critical concentration. Below this threshold (labelled as $c_1$ in Fig. 3d in our experiment), added electrolytes primarily improve the conductivity, thus enhancing the electrolysis efficiency. Based on our experiments, we determined that the threshold is approximately 0.025 M for both $HClO_4$ and $Na_2SO_4$, as shown in Fig. 3c. Above this threshold, the efficiency decreases with concentration, indicating that the adverse effects of bubble coalescence inhibition begin to outweigh the beneficial effects of enhanced conductivity. The electrolysis efficiency starts to increase after another concentration threshold, $c_2$ (~0.3 M for $HClO_4$ and ~0.5 M for $Na_2SO_4$), suggesting that the negative effects of coalescence inhibition are outweighed by the increased conductivity with the further increase in electrolyte concentration (Fig. 3b). The two-transition threshold, $c_1$ and $c_2$, aligns well with the concentration–coalescence inhibition effect of electrolytes[35,38,42], suggesting that coalescence inhibition has a more substantial effect on electrolysis efficiency than conductivity over a broad concentration range.

To better illustrate the impact of bubble-bubble interaction on electrolysis efficiency, we constructed diagrams for $HClO_4$ and $Na_2SO_4$ (Figs. 3e and f, respectively), using ΔE (defined as $E_{(0.5\ M\ H_2SO_4 + X\ M\ HClO_4/Na_2SO_4)} - E_{(0.5\ M\ H_2SO_4)}$) to compare changes in electrolysis efficiency with the added electrolytes (the detailed experimental data is provided in Figs.S4 and S5). Note that a negative ΔE observed at relatively high currents, e.g., -30 mA for $HClO_4$, indicates a clear reduction of electrolysis efficiency compared to that in 0.5 M $H_2SO_4$.

The diagrams clearly demonstrate the impact of bubble-bubble interactions on electrolysis efficiency. This effect becomes more pronounced at higher electrolysis currents (where bubble collision and coalescence frequency increase) and depends on the type and concentration of the added electrolyte. Notably, a competitive interplay is observed between conductivity enhancement and bubble coalescence inhibition when adding electrolytes. While increased conductivity generally enhances electrolysis efficiency, the inhibition of bubble coalescence at specific electrolyte concentrations (between thresholds $c_1$ and $c_2$) can offset these benefits.

**Explanations for how bubble coalescence improves electrolysis efficiency**

The findings collectively suggest that bubble-bubble interactions significantly influence electrolysis efficiency. It is essential to discuss the underlying mechanism. Analysis of the experimental data indicates that the key process is the continuous coalescence of newly detached bubbles with surface bubbles, which removes surface bubbles from the electrode at much smaller sizes and enhances mass and heat transfer at the electrolyte-electrode interface.

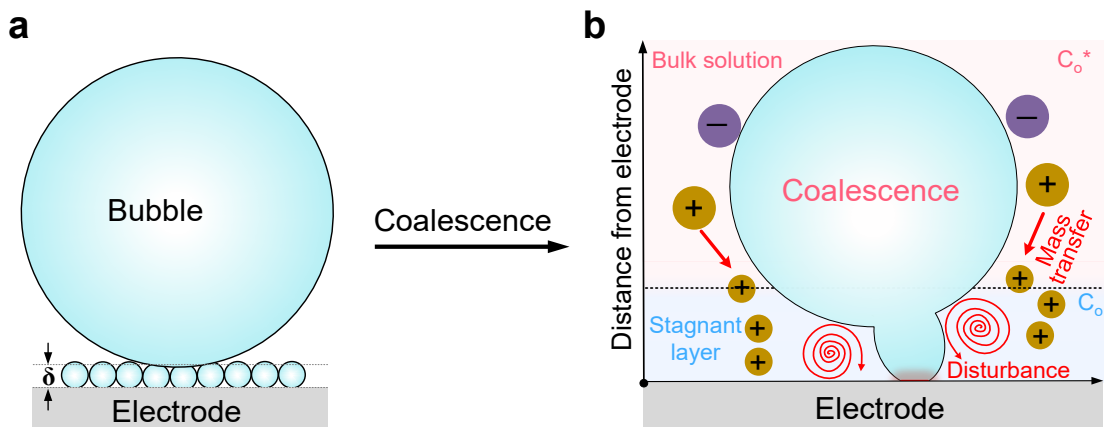

**Figure 4. | A discussion on how bubble coalescence improves electrolysis efficiency. a,** Schematic of the bubble detachment process where a just-detached large bubble continuously coalesces with the microbubble layer on the electrode surface, with a thickness δ of approximately 10 μm[43]. **b,** A schematic illustrating the intense disturbance caused by the merging of a surface bubble into the above larger one. $C_o$ represents the $H^+$ concentration of the solution in the stagnant layer, while $C_o*$ denotes the $H^+$ concentration of the bulk solution. This coalescing removes surface bubbles from the electrode at much smaller sizes and enhances mass and heat transfer at the electrolyte-electrode interface.

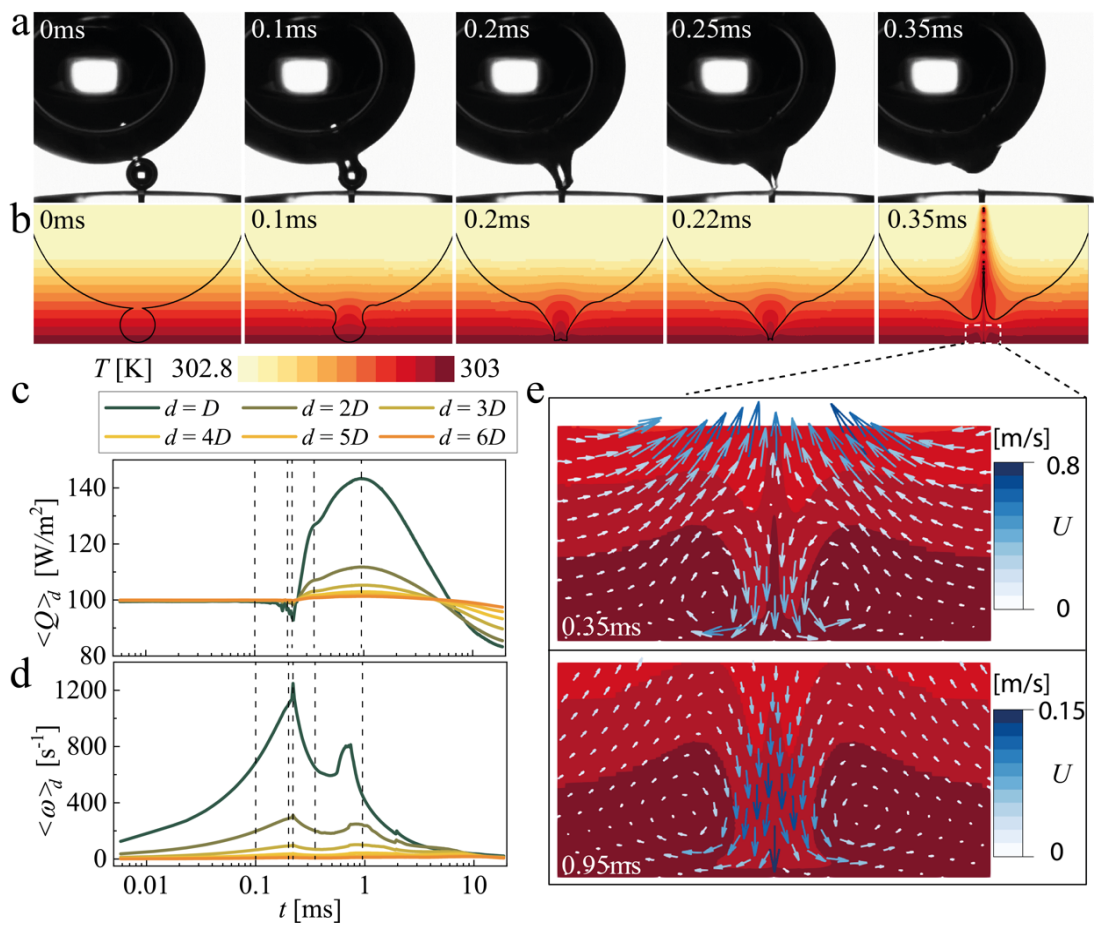

**Figure 5. | Bubble coalescence improves electrolysis efficiency via swirling hydrodynamics. a,** Image sequence depicting the coalescence of a large bubble (6.33$D$ in diameter), generated by a capillary (diameter $\Phi = 50$ μm), with a smaller surface bubble (diameter $D = 450$ μm) at the working electrode. **b,** Numerical solutions depicting the changes in bubble profile and temperature distribution during coalescence. The agreement in the evolving bubble profiles obtained from both simulations and experimental observations validates the accuracy of the numerical model. **c,** Heat transfer enhancement during coalescence, illustrated by the averaged heat flux across the interface of various surface domains ($d = D, 2D, 3D, 4D, 5D,$ and $6D$). **d,** Onset of strong vortices during bubble coalescence, illustrated by the volume-averaged vorticity of the electrolyte within the corresponding volume domains. **e,** Detailed visualization of flow filed after the violent pulling up of the bottom bubble upon coalescence.

First, the coalescence process helps release surface bubbles from the electrode at much smaller sizes (see Figure 5a and Supplementary video 4). In high-current electrolysis, a layer of surface bubbles approximately 10 μm[43] thick forms on the electrode (Fig. 4a and supplementary video 4). Surface bubbles larger than this layer thickness, δ, collide with the newly detached bubble above them, subsequently merging into it. This merging process displaces larger surrounding surface bubbles, allowing them to be absorbed into the detached bubble, which determines the bubble layer thickness—approximately 10 μm[27]. The size of the existing surface bubbles is much smaller than the detached bubbles observed in the low-current electrolysis with negligible bubble coalescence, with the latter measuring approximately 60-80 μm (see Fig. 2).

Furthermore, bubble coalescence creates intense agitation in the traditionally considered "stagnant" electrode-electrolyte interfacial region, thereby enhancing local mass and heat transfer. As shown in Fig. 5a, the surface bubble violently rises by approximately 100 μm within a brief timeframe of 0.25 to 0.3 ms upon coalescence, indicating a velocity on the order of ∼1m/s. (Note that the bubbles are generated by a glass capillary for better visualization; Supplementary Video 5 illustrates the same characteristics of electrolytic bubbles.) Conventionally, the electrode-electrolyte interfacial region is considered stagnant due to the no-slip boundary condition at the solid-liquid interface. Consequently, diffusive transport predominates in the near-surface region, limiting electrolysis efficiency [44]. However, the intense agitation caused by bubble coalescence disrupts this stagnant layer and introduces convective transport, thereby enhancing mass and heat transfer rates, as illustrated in Fig. 4b.

To quantitatively demonstrate the enhanced transport efficiency resulting from bubble coalescence, we numerically model the coalescence of a detached bubble (diameter $6.33D$) and a surface bubble ($D = 450$ μm in diameter) resting on a flat electrode surface at a constant temperature, using the open-source solver Basilisk[45]. Our axisymmetric model governs the flow field using the Navier-Stokes equations under the assumption of constant surface tension. We focus on the evolution of the thermal field through the advective diffusion equation, simplifying the simulation by omitting the species concentration fields. Since both temperature and mass fraction follow the same advective diffusion process[46], enhanced convection in the thermal field also implies improved species mixing, further boosting electrolysis efficiency near the electrode surface. Additional numerical settings based on the experiments are detailed in the supplementary material. The simulation was validated by comparing the characteristic evolution of bubble profiles from experiments (Fig. 5a) with simulations results (black solid lines in Fig. 5b).

The simulation results reveal intense agitation in the previously stagnant electrode-electrolyte interfacial region. As shown in Fig. 5b, the initially steady thermal boundary layer (color-coded) at the electrode surface is disrupted by bubble coalescence, which triggers a Rayleigh jet at 0.35 ms, forming micro-droplets within the bubble and a daughter bubble in the bulk[47]. Fig. 5c and 5d illustrate the evolution of the averaged thermal flux at the electrode surface, $\langle Q \rangle_d$, and the averaged vorticity in the electrolyte, $\langle \omega \rangle_d$, with the size of the averaging space color-coded (definitions in SI). Vertical dashed lines indicate key time points ($t =$ 0.1 ms, 0.2 ms, 0.22 ms, 0.35 ms, and 0.95 ms). A sharp increase in thermal flux is observed around 0.2 ms, coinciding with peak vorticity, and remains elevated for approximately 10 ms due to the localized convective motion. Snapshots at 0.35ms and 0.95ms highlight the disruption of the thermal boundary layer by the coalescence-induced vortex, with arrows indicating local velocity and bluish colour representing velocity magnitude.

The average heat flux across the electrode surface, previously covered by the bottom bubble, increased by about 40% due to coalescence. Notably, the impact on heat transfer lasts for about 10 ms—two orders of magnitude longer than the 0.2 ms coalescence process itself. These strong, long-lasting vortices are also similarly expected to greatly enhance mass transport at the electrolyte-electrode interface.

**Conclusion**

In this study, we highlight the crucial role of bubble coalescence in water electrolysis efficiency. For example, adding 0.3 M $HClO_4$, which inhibits coalescence, can reduce efficiency by more than 30%, despite improving conductivity. We identified two key mechanisms through which enhanced bubble coalescence improves efficiency: (1) while the coalescence process significantly increases the departure size of newly detached bubbles, it actually helps release the bubbles from the electrode at a much smaller size; (2) the coalescence process generates strong, long-lasting vortices that significantly enhance mass and heat transfer at the electrolyte-electrode interface.

This work provides several important insights. First, there is great potential to improve electrolysis efficiency by enhancing bubble coalescence in systems where it is inhibited, such as alkaline water/seawater electrolysis, the chlor-alkali process, and the Hall–Heroult process. Second, the findings from this work suggest new pathways to enhance electrolysis efficiency, such as through electrode design and electrolyte engineering, which can regulate bubble coalescence by adjusting bubble collision frequency and coalescence probability. Finally, since bubble-bubble interactions fundamentally govern bubbling behaviour, we believe this understanding applies not only to water electrolysis but also to other gas-evolving electrochemical systems.

**Data availability**

During the preparation of this work, the author(s) used ChatGPT to improve the language. After using this tool, the author(s) reviewed and edited the content as needed and take(s) full responsibility for the content of the publication. The data that support the findings of this study are available from the corresponding author upon request. Source data are provided with this paper.

**Acknowledgements** This work was financially supported by the National Natural Science Foundation of China (Grant.Nos.52025103,22178099,12102171), the Shanghai Pilot Program for Basic Research (Grant. No. 22TQ1400100-11）, the Special Project for Peak Carbon Dioxide Emissions-Carbon Neutrality (Grant.No.21DZ1207800) from the Shanghai Municipal Science and Technology Commission, and the Guang Dong Basic and Applied Basic Research Foundation (Grant Nos. 2024A1515010509, 2024A1515010614).

**Author contributions** All authors wrote and read the paper. All authors conceived and designed the experiments. T.W. Writing–original draft, Methodology, Investigation. B.L. Conceptualization, Writing – review & editing, Methodology, Investigation, Supervision. H.H.H. Writing – review & editing. F.Y. Review & editing, Methodology. Y.Z. Review & editing. H.S.T. Review & editing, Methodology. Q.Y. Review & editing, Funding acquisition, Supervision.

**Competing interests** The authors declare no competing interests

**References**

1. Shiva Kumar, S. & Lim, H. An overview of water electrolysis technologies for green hydrogen production. *Energy Rep.* **8**, 13793–13813 (2022).
2. Muradov, N. Z. & Veziroğlu, T. N. "Green" path from fossil-based to hydrogen economy: An overview of carbon-neutral technologies. *Int. J. Hydrog. Energy* **33**, 6804–6839 (2008).
3. Angulo, A., van der Linde, P., Gardeniers, H., Modestino, M. & Fernández Rivas, D. Influence of Bubbles on the Energy Conversion Efficiency of Electrochemical Reactors. *Joule* **4**, 555–579 (2020).
4. Zeradjanin, A. R., Narangoda, P., Spanos, I., Masa, J. & Schlögl, R. How to minimise destabilising effect of gas bubbles on water splitting electrocatalysts? *Curr. Opin. Electrochem.* **30**, 100797 (2021).
5. Zhan, S. *et al.* Experimental investigation on bubble growth and detachment characteristics on vertical microelectrode surface under electrode-normal magnetic field in water electrolysis. *Int. J. Hydrog. Energy* **46**, 36640–36651 (2021).
6. Zhang, D. & Zeng, K. Evaluating the Behavior of Electrolytic Gas Bubbles and Their Effect on the Cell Voltage in Alkaline Water Electrolysis. *Ind. Eng. Chem. Res.* **51**, 13825–13832 (2012).
7. Koza, J. A. *et al.* Hydrogen evolution under the influence of a magnetic field. *Electrochimica Acta* **56**, 2665–2675 (2011).
8. Yang, G. *et al.* Building Electron/Proton Nanohighways for Full Utilization of Water Splitting Catalysts. *Adv. Energy Mater.* **10**, 1903871 (2020).
9. Zeng, T. *et al.* Manageable Bubble Release Through 3D Printed Microcapillary for Highly Efficient Overall Water Splitting. *Adv. Sci.* **10**, 2207495 (2023).
10. Zeradjanin, A. R. *et al.* Rational design of the electrode morphology for oxygen evolution – enhancing the performance for catalytic water oxidation. *RSC Adv.* **4**, 9579–9587 (2014).
11. Lin, C. *et al.* In-situ reconstructed Ru atom array on α-MnO2 with enhanced performance for acidic water oxidation. *Nat. Catal.* **4**, 1012–1023 (2021).
12. Fogaça, W., Ikeda, H., Misumi, R., Kuroda, Y. & Mitsushima, S. Enhancement of oxygen evolution reaction in alkaline water electrolysis by Lorentz forces generated by an external magnetic field. *Int. J. Hydrog. Energy* **61**, 1274–1281 (2024).
13. Gatard, V., Deseure, J. & Chatenet, M. Use of magnetic fields in electrochemistry: A selected review. *Curr. Opin. Electrochem.* **23**, 96–105 (2020).
14. Lin, M.-Y. & Hourng, L.-W. Ultrasonic wave field effects on hydrogen production by water electrolysis. *J. Chin. Inst. Eng.* **37**, 1080–1089 (2014).

15. Vogt, H. & Balzer, R. J. The bubble coverage of gas-evolving electrodes in stagnant electrolytes. *Electrochimica Acta* **50**, 2073–2079 (2005).
16. Yang, X., Karnbach, F., Uhlemann, M., Odenbach, S. & Eckert, K. Dynamics of Single Hydrogen Bubbles at a Platinum Microelectrode. *Langmuir* **31**, 8184–8193 (2015).
17. Hossain, S. S., Bashkatov, A., Yang, X., Mutschke, G. & Eckert, K. Force balance of hydrogen bubbles growing and oscillating on a microelectrode. *Phys. Rev. E* **106**, 035105 (2022).
18. Liu, B., Manica, R., Xu, Z. & Liu, Q. The boundary condition at the air-liquid interface and its effect on film drainage between colliding bubbles. *Curr. Opin. Colloid Interface Sci.* **50**, (2020).
19. Liu, B. *et al.* Dynamic Interaction between a Millimeter-Sized Bubble and Surface Microbubbles in Water. *Langmuir* **34**, 11667–11675 (2018).
20. Liu, B. *et al.* Nanoscale Transport during Liquid Film Thinning Inhibits Bubble Coalescing Behavior in Electrolyte Solutions. *Phys. Rev. Lett.* **131**, 104003 (2023).
21. Enright, R. *et al.* How Coalescing Droplets Jump. *ACS Nano* **8**, 10352–10362 (2014).
22. Lv, P. *et al.* Self-Propelled Detachment upon Coalescence of Surface Bubbles. *Phys. Rev. Lett.* **127**, 235501 (2021).
23. Raza, Md. Q. *et al.* Coalescence-induced jumping of bubbles in shear flow in microgravity. *Phys. Fluids* **35**, 023333 (2023).
24. Zhao, P., Hu, Z., Cheng, P., Huang, R. & Gong, S. Coalescence-Induced Bubble Departure: Effects of Dynamic Contact Angles. *Langmuir* **38**, 10558–10567 (2022).
25. Iwata, R. *et al.* How Coalescing Bubbles Depart from a Wall. *Langmuir* **38**, 4371–4377 (2022).
26. Bashkatov, A. *et al.* Performance Enhancement of Electrocatalytic Hydrogen Evolution through Coalescence-Induced Bubble Dynamics. *J. Am. Chem. Soc.* **146**, 10177–10186 (2024).
27. Bashkatov, A. *et al.* On the growth regimes of hydrogen bubbles at microelectrodes. *Phys. Chem. Chem. Phys.* (2022) doi:10.1039/D2CP02092K.
28. Park, S., Lohse, D., Krug, D. & Koper, M. T. M. Electrolyte design for the manipulation of gas bubble detachment during hydrogen evolution reaction. *Electrochimica Acta* 144084 (2024) doi:10.1016/j.electacta.2024.144084.
29. Han, Y., Bashkatov, A., Huang, M., Eckert, K. & Mutschke, G. Impact of tracer particles on the electrolytic growth of hydrogen bubbles. *Phys. Fluids* **36**, 012107 (2024).
30. Liu, Y. *et al.* Effects of magnetic field on water electrolysis using foam electrodes. *Int. J.*

*Hydrog. Energy* **44**, 1352–1358 (2019).

31. Brickwedde, L. H. Properties of aqueous solutions of perchloric acid. *J. Res. Natl. Bur. Stand.* **42**, 309 (1949).

32. Blandamer, M. J., Franks, F., Haywood, K. H. & Tory, A. C. Effect of Added Solutes on the Size of Hydrogen Bubbles liberated from a Cathodic Wire in Aqueous Solution. *Nature* **216**, 783–784 (1967).

33. Yang, X., Karnbach, F., Uhlemann, M., Odenbach, S. & Eckert, K. Dynamics of Single Hydrogen Bubbles at a Platinum Microelectrode. *Langmuir* **31**, 8184–8193 (2015).

34. Craig, V. S. J., Ninham, B. W. & Pashley, R. M. Effect of electrolytes on bubble coalescence. *Nature* **364**, 317–319 (1993).

35. Henry, C. L., Dalton, C. N., Scruton, L. & Craig, V. S. J. Ion-Specific Coalescence of Bubbles in Mixed Electrolyte Solutions. *J. Phys. Chem. C* **111**, 1015–1023 (2007).

36. Firouzi, M., Howes, T. & Nguyen, A. V. A quantitative review of the transition salt concentration for inhibiting bubble coalescence. *Adv. Colloid Interface Sci.* **222**, 305–318 (2015).

37. Craig, V. S. J. Bubble coalescence and specific-ion effects. *Curr. Opin. Colloid Interface Sci.* **9**, 178–184 (2004).

38. Craig, V. S. J. & Henry, C. L. Specific Ion Effects at the Air?Water Interface: Experimental Studies. in *Specific Ion Effects* 191–214 (WORLD SCIENTIFIC, 2009). doi:10.1142/9789814271585_0007.

39. Kristof, P. & Pritzker, M. Effect of electrolyte composition on the dynamics of hydrogen gas bubble evolution at copper microelectrodes. *J. Appl. Electrochem.* **27**, 255–265 (1997).

40. Mukouyama, Y., Kikuchi, M. & Okamoto, H. Appearance of new potential oscillation during hydrogen evolution reaction by addition of Na2SO4 and K2SO4. *J. Electroanal. Chem.* **617**, 179–184 (2008).

41. Pashley, R. M. & Craig, V. S. J. Effects of Electrolytes on Bubble Coalescence. *Langmuir* **13**, 4772–4774 (1997).

42. Lessard, R. R. & Zieminski, S. A. Bubble Coalescence and Gas Transfer in Aqueous Electrolytic Solutions. *Ind. Eng. Chem. Fundam.* **10**, 260–269 (1971).

43. Bashkatov, A., Hossain, S. S., Yang, X., Mutschke, G. & Eckert, K. Oscillating Hydrogen Bubbles at Pt Microelectrodes. *Phys. Rev. Lett.* **123**, 214503 (2019).

44. Khalighi, F., Vreman, A. W., Tang, Y. & Deen, N. G. Effects of the boundary conditions at the gas-liquid interface on single hydrogen bubble growth in alkaline water electrolysi. *Chem. Eng. Sci.* **301**, 120666 (2025).

45. Popinet, S. An accurate adaptive solver for surface-tension-driven interfacial flows. *J. Comput. Phys.* **228**, 5838–5866 (2009).
46. Bergman, T. L. & Lavine, A. S. *Fundamentals of Heat and Mass Transfer*. (John Wiley & Sons, Hoboken, NJ, 2017).
47. Tufaile, A. & Sartorelli, J. C. Bubble and spherical air shell formation dynamics. *Phys. Rev. E* **66**, 056204 (2002).